\documentclass[12pt]{iopart}

\begin{document}

\title{Bulk viscosity in nuclear and quark matter: a short review}

\author{Hui Dong$^1$, Nan Su$^2$, Qun Wang$^1$}

\address{$^1$Department of Modern Physics, University of Science and
Technology of China, Anhui 230026, China\\
$^2$Frankfurt Institute for Advanced Studies, D-60438 Frankfurt
am Main, Germany}

\begin{abstract}
The history and recent progresses in the study of bulk viscosity 
in nuclear and quark matter are reviewed. 
The constraints from baryon number conservation and electric neutrality 
in quark matter on particle densities and fluid velocity 
divergences are discussed.

\end{abstract}

%\maketitle

The time scale for damping of the vibration 
and of the gravitational radiation driven instability in 
neutron stars is crucial to the stellar stability, 
which is controlled by the shear and bulk viscosities. 
The bulk viscosity originated from the re-establishment of 
chemical equilibrium is important in some circumstances. 
In this short note, we will give an overview on 
the history and recent progresses in the study of bulk viscosity
in nuclear and quark matter.

\section{Nuclear matter}

Sawyer~\cite{Sawyer:1989dp} pointed out that 
at temperatures higher than $10^9$ K the bulk viscosity for 
neutron matter is much larger than the shear viscosity. 
This is determined by opposite temperature behavior of the bulk and shear 
viscosities that the bulk viscosity increases while the shear decreases 
with growing temperature. Since then, the bulk viscosities 
in nuclear matter have been widely investigated.

%\section{$npe\mathrm{(}\mu\mathrm{)}$ matter}

{\em $npe\mathrm{(}\mu\mathrm{)}$ matter.}
Low density neutron star matter is composed of 
neutrons $n$ and a small admixture of protons $p$, 
electrons $e$, and possibly muons $\mu$.  
The bulk viscosity is mainly determined by the modified Urca processes,
\begin{equation}
N + n  \rightarrow  N  +  p  +  l  +  \overline{\nu}_l, \;\;\;\;
N  +  p  + l  \rightarrow  N  +  n +  \nu _l,
\label{eq:modurca}
\end{equation}
where $N$ denotes the spectator nucleon for 
energy and momentum conservation, $l$ the electron or muon, 
and $\nu _l$ the associated neutrino. In late 1960s, Finzi and Wolf 
analyzed the damping of neutron star pulsations 
via modified Urca processes in $npe$ matter \cite{Finzi:1968}, followed by 
Sawyer~\cite{Sawyer:1989dp} and Haensel \textit{et al.}~\cite{Haensel:2001mw} 
who investigated the bulk viscosity for $npe$ and $npe\mu$ matter respectively.  
The relaxation time for the modified Urca processes 
is of order $\tau \propto T^{-6}$.

At densities of a few times normal nuclear matter density, 
the direct Urca processes,
\begin{equation}
n \rightarrow  p  +  l  +  \overline{\nu}_l, 
\ \ \ \ \ \ p  +  l  \rightarrow  n  +  \nu _l,
\label{eq:directurca}
\end{equation}
may also be allowed provided the proton fraction
exceeds the Urca limit of about 1/9 \cite{Lattimer:1991ib}. 
The bulk viscosity was calculated for the direct Urca processes 
by Haensel and Schaeffer~\cite{Haensel:1992} in $npe$ matter and
by Haensel \textit{et al.}~\cite{Haensel:2000vz} in $npe\mu$ matter. 
A weaker temperature dependence for the relaxation time, 
$\tau \propto T^{-4}$, is found due to a smaller number of particles 
involved. Consequently the bulk viscosity for the direct Urca 
is about 4-6 orders of magnitude larger than that from the modified Urca 
at typical neutron star temperatures, $T\sim 10^9$-$10^{10}$ K. 
The large difference occurs at low temperatures. 

There may be superfluids in neutron star matter due to 
attractive part of baryon-baryon potential, 
for reviews, see e.g., Refs. \cite{Yakovlev:1999sk,Lombardo:2000ec}. 
The bulk viscosity with superfluidity for the direct and modified Urca processes 
in $npe\mu$ matter was investigated by Haensel 
\textit{et al.}~\cite{Haensel:2001mw,Haensel:2000vz}, 
which is substantially suppressed compared to normal state matter.

%\section{Hyperon matter}

%Strong interaction
%processes do not play a role, because the strong interaction
%equilibrium is reached so fast that these processes can be
%considered to be in thermal equilibrium compared to the typical
%pulsation time-scales of $10^{-4}-10^{-3} \ \textrm{s}$. 

{\em Hyperon matter.}
Hyperons may appear in the neutron star core 
\cite{Knorren:1995ds,Balberg:1997yw,Baldo:1998hd,Glendenning:1997ak}.  
With increasing densities, $\Sigma ^{-}$ and $\Lambda $ hyperons appear, 
followed by $\Xi ^0$, $\Xi ^{-}$ and $\Sigma ^{+}$. 
Most authors only considered $\Sigma ^{-}$ and $\Lambda $ hyperons which 
have the lowest threshold densities. 
Weak nonleptonic hyperon processes such as 
\begin{eqnarray}
%$nnp\Sigma^{-}$ process,
n  +  n   \leftrightarrow  p  +  \Sigma ^{-},\;\;\;
%the $pnp\Lambda$ process,
n  +  p  \leftrightarrow  p  +  \Lambda,\;\;\;
%the $nnn\Lambda$ process,
n  +  n  \leftrightarrow  n  +  \Lambda 
\label{nnnL} 
\end{eqnarray}
dominate the bulk viscosity, while their direct Urca processes 
\cite{Prakash:1992} have negligible contributions although 
they are comparable to nucleons' Urca. 
The relaxation time from processes (\ref{nnnL}) 
is of order $T^{-2}$. 

Langer and Cameron introduced the subject of bulk viscosity 
in hyperonic matter by estimating the damping of neutron star vibrations 
\cite{Langer:1969}. Jones carried out the 
semi-quantitative calculation of bulk
viscosity in hyperonic matter \cite{Jones:1971,Jones:2001ie}, where 
the weak nonleptonic process $nn\leftrightarrow p\Sigma ^{-}$ 
through $W$-exchange was calculated.  
Recently, Jones~\cite{Jones:2001ya} studied the reaction 
$nn\leftrightarrow n\Lambda$, the dominant channel for $\Lambda$ production 
in hypernuclei experiments in general, which cannot be mediated by single $W$ boson.  
The rate of the reaction $nn\leftrightarrow n\Lambda$ from the data 
is found to be several orders of magnitude larger than $nn\leftrightarrow p\Sigma ^{-}$ 
via the $W$-exchange. All of these works are the order-of-magnitude estimates. 

The various weak nonleptonic hyperon processes have
been recalculated by several authors with a modern equation of state. 
Haensel {\it et al.}~\cite{Haensel:2001em} studied the 
$nn\leftrightarrow p\Sigma ^{-}$ process within the nonrelativistic limit, 
they found the bulk viscosity to be several orders of 
magnitude larger than that of the direct and modified Urca processes. 
Lindblom and Owen~\cite{Lindblom:2001hd} computed the contribution of the
$np\leftrightarrow p\Lambda$ in addition to the $nn\leftrightarrow p\Sigma ^{-}$ process. 
The superfluidity case had also been studied in both works.

Most of the weak hadronic processes 
involved in the $nn\leftrightarrow n\Lambda$ process 
also contribute to the $nn\leftrightarrow p\Sigma ^-$ one, 
so the rates of these two processes should be of the same order, 
opposite to Jones' estimate. 
Dalen and Dieperink~\cite{vanDalen:2003uy} studied the bulk viscosity 
from all three processes in (\ref{nnnL}) using one-pion-exchange (OPE), which 
can well describe the rates of these processes in hypernuclei. 
Their results showed that the bulk viscosity in the OPE picture is about 
1-2 orders of magnitude smaller than that with $W$-exchange.

Recent hypernuclei data showed the potentials for $\Lambda$ and $\Xi$ are 
attractive but that for $\Sigma$ is repulsive in normal nuclear matter 
\cite{Friedman:1994hx,Batty:1997zp}. 
Chatterjee and Bandyopadhyay~\cite{Chatterjee:2006hy} took this fact 
into account and found the disappearance of $\Sigma$ in nuclear matter. 
They calculated the bulk viscosity from $np\leftrightarrow p\Lambda$. 
Their results showed that the bulk viscosity incorporating the hyperon
potentials implied by the data is larger than that without them.

%%%%%%%%%%%%%%%%%%%%%%%%%%%% Su Nan

\section{Quark matter}

When the baryon density is above 5-10 times the normal nuclear matter 
density, the deconfinement phase transition will take place 
marked by the formation of quark matter. 
Initially there are only $u$ and $d$ quarks, 
as the density grows the $s$ quarks will appear.  
Quark matter with $u$, $d$, and $s$ quarks can be self-bound 
and be called strange quark matter \cite{Collins:1974ky,Witten:1984rs}. 
The star made of strange quark matter is called quark star 
or strange star \cite{Alcock:1986hz,Haensel:1986qb}.  

{\em Nonleptonic processes.}
The importance of dissipation due to the nonleptonic reaction,  
\begin{equation}
\label{nonleptonic}
s+u\leftrightarrow u+d.
\end{equation}
was first observed by Wang and Lu \cite{Wang:1985tg}. 
They showed that stellar pulsations would be strongly damped 
in quark matter. Saywer \cite{Sawyer:1989uy} then formulated 
the damping in terms of the bulk viscosity based the linear 
expansion of the reaction rate of (\ref{nonleptonic}) in 
$\delta\mu=\mu_s-\mu_d$, where $\mu_i$ with 
$i=s,d$ are the quark chemical potentials. 
In the temperature range characteristic of young neutron stars, the
bulk viscosity arising from the reaction (\ref{nonleptonic}) 
is orders of magnitude larger than that for normal nuclear matter.
%Thus the secular instability of gravitational wave radiation 
%should be more suppressed for quark stars with very short 
%damping time for pulsations. 
However the above linear assumption is not proper at low temperatures 
($T\ll\delta\mu$), where the rate is proportional 
to $\delta\mu^3$. Madsen~\cite{Madsen:1992sx} 
showed that this nonlinearity effect
leads to much larger the bulk viscosity than previously
assumed. Note that strong interactions 
can also influence the rate of the reaction (\ref{nonleptonic}) 
and the bulk viscosity of strange quark matter significantly \cite{Dai:1996fe}. 
At low temperatures, the viscosity
is strongly suppressed, while at high temperatures it is slightly
enhanced. The above calculations are based on MIT bag model where 
all quark masses are taken to be constants. Alternatively 
quark massess can be assumed to depend on baryon density 
\cite{Fowler:1981rp}. By employing this density dependent quark
model, the bulk viscosity at low temperatures and high relative
perturbations increases 2-3 orders of magnitude, while at low
perturbations the enhancement is 1-2 orders of magnitude 
compared to the results obtained in other approaches 
\cite{Anand:1999bj}. 
%In analogy to the Bardeen-Cooper-Schrieffer
%theory of superconductivity \cite{Bardeen:1957mv},
Sufficiently dense and cold quark matter is expected to be a 
color superconductor \cite{Bailin:1983bm}.  
A recent interest in the study of the bulk viscosity 
in color superconducting phases is growing, for example, 
the bulk viscosity in the 2-flavors 
color superconducting phase (2SC) from the reaction (\ref{nonleptonic}) 
has been computed by Alford and Schmitt \cite{Alford:2006gy}. 
The bulk viscosity due to kaons in color-flavor-locked (CFL)
phase has also been calculated \cite{Alford:2007rw}.

%Wang and Lu (1984), Sawyer (1989), Madsen (1992), Dai and Lu (1996),
%Anand etc (2000), Alford and Schmitt (2006)

%\section{Leptonic processes}

{\em Leptonic processes.}
Due to phase space restrictions \cite{Wang:2006tg,Wang:2006xf} 
the rates of the leptonic processes,
\begin{equation}
\label{urca}
u+e\rightarrow q+\nu_e,\;\;\;
q\rightarrow u+e+\overline{\nu}_e,\;\;\; (q=d,s)
\end{equation}
are much smaller than nonleptonic ones at low temperatures. 
These Urca processes are the most efficient way to cool 
the neutron stars. As shown by Anand {\it et al.} \cite{Anand:1999bj}, 
when the temperature increases the contribution of leptonic processes to 
the bulk viscosity will exceed that from nonleptonic ones. 
So more careful studies are needed to
explore the dependence of bulk viscosity on the broader ranges of
temperature. The energy emissivity in Urca processes 
in various color superconducting phases have been studied by several groups 
\cite{Alford:2004zr,Jaikumar:2005hy,Schmitt:2005wg,Anglani:2006br}. 
The bulk viscosity from the Urca processes for 
$d$ quarks in a spin-one color superconductor has been 
calculated by Sa'd, Shovkovy and Rischke \cite{Sa'd:2006qv}.

%Anand etc (2000), Sa'd etc (2006)

%\section{More realistic case}

\section{Baryon number conservation and enforced charge neutrality 
for bulk viscosity in quark matter}

A more realistic case for the bulk viscosity in quark matter 
is to take both of nonleptonic and leptonic precesses into account. 
There must be electrons to compensate the net positive charge in 
a three-flavor quark system because the mass of $s$ quarks is much larger 
than those of light quarks. The bulk viscosity should be 
calculated for charge neutral quark matter. 
In nonleptonic and leptonic reactions, the electron number 
and flavors are not conserved, while the baryon number and electric charge
are conserved. From the continuity equations, baryon number 
conservation and charge neutrality, we can derive the following 
constraints for particle number densities 
and velocity divergences \cite{Dong:2007mb}, 
\begin{equation}
n_{B}\nabla\cdot\mathbf{v}_{B} =  
\sum_{i=u,d,s}\frac{1}{3}n_{i}\nabla\cdot\mathbf{v}_{i}\,,
\,\,\,\,\,\,\,\,\,\,
n_{e}\nabla\cdot\mathbf{v}_{e}  =  
\sum_{i=u,d,s}Q_{i}n_{i}\nabla\cdot\mathbf{v}_{i}\,, 
\label{eq:cont3}
\end{equation}
where $n_j$, $Q_j$ and $\mathbf{v}_{j}$ 
are number densities, electric charges and fluid velocities 
for particle $j=B,e,u,d,s$ respectively. The indices $B$ and $e$  
denote baryons and electrons. 
Here the first equation is from baryon number conservation and
the second one from charge neutrality. 
All previous treatments in the literature used the conditions 
$n_{B}dX_{i}/dt=J_{i}$ to determine the bulk viscosity, 
where $X_{i}\equiv n_i/n_B$ are fractions of baryon number 
for particle $i=u,d,s,e$ and $J_{i}$ their sources. 
Using $dn_B/dt=-n_B\nabla\cdot\mathbf{v}_{B}$, we can get
$n_{B}dX_{i}/dt=dn_{i}/dt-X_{i}dn_{B}/dt
=dn_{i}/dt+n_i\nabla\cdot\mathbf{v}_{B}=J_i$.
Comparing it with the continuity equation for 
particle $i$, $dn_{i}/dt+n_i\nabla\cdot\mathbf{v}_{i}=J_i$, 
we obtain $\nabla\cdot\mathbf{v}_{B}=\nabla\cdot\mathbf{v}_{i}$,  
which obey Eq. (\ref{eq:cont3}) obviously.  
Hence the usage of $n_{B}dX_{i}/dt=J_{i}$ 
in literature implies the unique value of velocity divergences 
for all particle species. However, considering the special 
role of strange quarks due to their large mass, we propose 
a new possibility that the velocity divergence for $s$ quarks 
is different from those of light quarks and electrons, which 
corresponds to a new oscillation pattern for the bulk viscosity. 
As an extreme case, we assume $\nabla\cdot\mathbf{v}_{s}=0$ and 
$\nabla\cdot\mathbf{v}_{u}=\nabla\cdot\mathbf{v}_{d}$.
From Eq. (\ref{eq:cont3}), we derive $\nabla\cdot\mathbf{v}_{e} 
=\frac{n_{B}(2n_{u}-n_{d})}{n_{e}(n_{u}+n_{d})}\nabla\cdot\mathbf{v}_{B}$ 
and $\nabla\cdot\mathbf{v}_{u,d}
=\frac{3n_{B}}{n_{u}+n_{d}}\nabla\cdot\mathbf{v}_{B}$.    
We can use $\delta n_B$, $\delta n_e$ and $\delta n_s$ 
as independent variables for the density oscillation. 
Then $\delta n_{u,d}$ and $\delta P$ can be expressed 
in terms of these independent variables. 
In order to close the system of equations for the bulk viscosity, 
we need two additional inputs, e.g. 
$d\delta n_e/dt+n_e\nabla\cdot\mathbf{v}_{e}$ and 
$d\delta n_s/dt$ (note that $\nabla\cdot\mathbf{v}_{s}=0$), 
given by the reaction rates which can also be expressed in terms of 
independent variables. The former can be written as 
$d\delta n_e/dt+n_e\nabla\cdot\mathbf{v}_{e}
=d\delta n_e/dt-(2n_u-n_d)/(n_u+n_d)d\delta n_B/dt$. 
Then the bulk viscosity is determined from the definition 
$\delta P=-\zeta \nabla\cdot\mathbf{v}_{B}$. 
The results are shown in Fig. 3 of Ref. \cite{Dong:2007mb}. 
The values of the bulk viscosity in this solution is 1-2 orders of 
magnitude larger than the conventional ones. 

Note that the assumption $\nabla\cdot\mathbf{v}_{s}=0$ reflects 
the consideration that the $s$ quarks are much heavier 
and may respond to the density oscillation more reluctantly 
than light particles. If the strange quark mass is small, 
one can investigate many other solutions which are close to 
the conventional solution 
$\nabla\cdot\mathbf{v}_{B}=\nabla\cdot\mathbf{v}_{i}$, 
for example, one can assume velocity divergences of 
particles deviate from that of baryons by a small amount. 

\textbf{Acknowledgement}. Q.W. acknowledges
support in part by the startup grant from University of Science and
Technology of China (USTC) in association with \emph{100 talents}
project of Chinese Academy of Sciences (CAS) and by National Natural
Science Foundation of China (NSFC) under the grant 10675109.

\end{document}